\documentstyle[aps]{revtex}

\begin{document}
\author{Yang Lei, Yang Kongqing}
\address{({\it Department of physics, Lanzhou University, Gansu 730000, China)}}
\title{Singularities in the Boussinesq equation and in the generalized KdV equation}
\date{Nov, 1th, 1999}
\maketitle

\begin{abstract}
In this paper, two kinds of analytic singular solutions (finite-time and
infinite-time singular solutions) of two classical wave equations (the
Boussinesq equation and a generalized KdV equation) are obtained by means of
the improved homogeneous balance method (HB method) and a nonlinear
transformation. The solutions show that special singular wave patterns exist
in the classical models of shallow water wave problem.

\end{abstract}

\section{Introduction}

All kinds of wave patterns [1], from linear systems to nonlinear systems,
contribute to understanding complex wave phenomena. Generally, the wave
patterns are symmetric resulting from the fascinating effects of
instability. When the instabilities of a nonlinear system are violent, the
energy may focus into a spike at a point where special wave patterns
(singularity in the nonlinear system) appear; the special wave patterns are
local divergences in the amplitude or gradients of some physical field. As
is known, a lot of spatio-temporal systems exhibit this phenomena, {\it e.g.}%
, in the turbulence problem it is observed in Ref [2] that large
fluctuations in the derivatives occur and the dissipation field appears
multifractal (a set of nested singularities); in the surface wave problem,
the singularity formation has been observed and studied [3,4]; nonlinear
optical systems exhibit self-focusing effects, which may lead to the
collapse of the optical power density into local divergences which may have
important consequences on the integrity of optical fibers and laser systems; 
{\it etc.}

For understanding singularities of nonlinear physical systems, it is helpful
to study singular solutions of partial differential equations (PDEs) that
model nonlinear physical systems, and many papers have been written on the
problem; for example, near-singular solutions of the Navier-Stokes equations
and singular solutions of the Euler equations [5]; the possibility of
singularities arising in Burger's equation [6]; some methods were presented
for the investigation of the singularity formation in the NLS equation [7];
near-singular solutions of the complex Ginzberg-Landau equation from the
view point that the NLS equation is the conservation limit of the complex
Ginzberg-Landau equations [8]; {\it etc}. However, most methods used in
studying singular solutions are based on perturbation techniques or the
singular point analysis. Also, the explicit forms of singular solution are
seldom discussed, except for rational solutions. In this paper, we try to
obtain the analytic singular solutions of nonlinear PDEs by two direct
methods: an improved HB method [9] and the invariant-B\"{a}cklund
transformation [10] based on a special nonlinear transformation.

The homogeneous balance method [11] has shown its efficiency in finding
analytic solitary wave solutions of many PDEs. Its essence can be presented
as follows: the nonlinear PDE\ is given by 
\begin{equation}
P(u,u_x,u_t,u_{xx},u_{xt},u_{tt},\cdots \cdots )=0.  \label{1}
\end{equation}
Supposing the solution of equation (1) is of the form 
\begin{equation}
u=\sum_{i=0}^Na_if^{(i)}(\varphi (x,t)),  \label{2}
\end{equation}
where $i$ is an integer, $a_i$ are constant coefficients, $f(\varphi (x,t))$
is a function of $\varphi (x,t)$, $\varphi (x,t)$ is a function of $x$ and $%
t $, and superscript $(i)$ represents the derivative index. According to the
assumption of homogeneous balance, in the PDE (1), the nonlinear terms and
the highest order partial derivative terms ought to be partially balanced.
Then $N$ is obtained, the expression of $f(\varphi (x,t))$ and the relation
of $f^{(i)}\cdot f^{(j)}$ and $f^{(i+j)}$ can be derived. Assuming $\varphi
(x,t)=b\ln (1+e^{\alpha x+\beta t+\gamma })$ and substituting formula (2)
into equation (1), after deciding coefficients $a_i,b,\alpha ,\beta ,\gamma $%
, the solitary ware solution of the nonlinear PDE is obtained. Recently an
improved HB method [9] has been reported: after balancing the nonlinear and
the highest order partial derivative terms, substituting $f(\varphi (x,t))$
and the relation of $f^{(i)}\cdot f^{(j)}$ and $f^{(i+j)}$ into equation
(1), we get

\[
F(f^{\prime },f^{\prime \prime },\ldots \varphi _x,\varphi _{xx},\ldots
\varphi _t,\varphi _{tt},\ldots \varphi _{xt},\ldots )=0, 
\]
where $F$ is a function of $f^{\prime },f^{\prime \prime },\ldots \varphi
_x,\varphi _{xx},\ldots \varphi _t,\varphi _{tt},\ldots \varphi _{xt},\ldots 
$ Obviously, $F$ is a linear polynomial of $f^{\prime },f^{\prime \prime
},\ldots $ Setting the coefficients of $f^{\prime },f^{\prime \prime
},\ldots $ to zero yields a set of partial differential equations of $%
\varphi (x,t)$. Taking note of the terms of $f^{\prime }$, their forms must
be $f^{^{\prime }}\varphi _{x_1,x_2,\ldots x_p,t_1\ldots t_q}$. So the
coefficient of $f^{^{\prime }}$ in $F(f^{\prime },f^{\prime \prime },\ldots
\varphi _x,\varphi _{xx},\ldots \varphi _t,\varphi _{tt},\ldots \varphi
_{xt},\ldots )=0$ is a linear polynomial $\sum\limits_{i=1}^ka_i\varphi
_{x_1,x_2,\ldots x_{pi},t_1\ldots t_{qi}}$, and then the set of partial
differential equations has an important feature, {\it i.e.} the equation
from the coefficients of $f^{\prime }$ is a linear PDE of $\varphi (x,t)$,
whilst the other equations can be regarded as constraint conditions to the
linear equation. Thus solving the set of PDEs reduces to solving the linear
PDE with some constraint conditions, and then the analytic solution
(including travelling and non-travelling wave solutions) of nonlinear PDEs
can be obtained by the improved HB method.

As an example, the Boussinesq equation and the KdV equation (fundamental
models of the shallow water wave problem) are discussed. Analytic singular
solutions of the equations are obtained by the improved homogeneous balance
(HB) method. An particular, a finite-time singular solution of the
Boussinesq equation is obtained in this paper, which was produced from a
non-singular physical field in the process of time evolution. The
finite-time singular solution is relevant to a conjecture---in hydrodynamics
many authors have shown that singularities of fluid motion can be formed in
finite time, and some simplified models of fluid motion have been presented
to demonstrate the conjecture [12]. Here the Boussinesq equation gives
another simplified model of fluid flow to do the same job, since this
equation is a fundamental model of the shallow water wave problem.

The B\"{a}cklund transformations were developed in the 1880s for use in the
related theories of differential geometry and differential equations.
Afterward, the relationships between the B\"{a}cklund transformations and
the inverse scattering transform [13] (IST) or the bilinear form [14] were
presented, and the B\"{a}cklund transformations have been used to find
analytic solutions of nonlinear PDEs. In this paper, based on a special
nonlinear transformation, the invariant-B\"{a}cklund transformation of a
generalized KdV equation is obtained and the analytic singular solution of
it is obtained.

This paper is organized as follows. In Sec II, an analytic singular solution
of the Boussinesq equation is obtained by the improved HB method, which
shows that the finite-time and the infinite-time singularity both exist in
the Boussinesq equation. In Sec III, an analytic singular solution of the
KdV equation is obtained by the improved HB method and an analytic periodic
singular solution in space of the generalized KdV equation is obtained by
the invariant-B\"{a}cklund transformation. In Sec IV, the main results are
given.

\section{Boussinesq equation}

The Boussinesq equation [15] is a classical model of long wavelength
hydrodynamic waves and other physical systems and is written in the form

\begin{equation}
u_{tt}-u_{xx}-a(u^2)_{xx}+bu_{xxxx}=0,  \label{3}
\end{equation}
where $a,b$ are real constants. It can be derived from the incompressible
fluid equations

\begin{equation}
\begin{array}{l}
\partial _tu=-\frac 1\rho \partial _xp^{\prime }, \\ 
\partial _tw=-\frac 1\rho \partial _zp^{\prime }, \\ 
\partial _xu+\partial _zw=0,
\end{array}
\label{4}
\end{equation}
where $u$, $w$ are velocities in the $x$, $z$ directions, $p^{\prime }$ is
the relative pressure. Eq. (3) can be derived from the Toda Lattice also.
The Boussinesq equation (for wave propagation in both directions) describes
wave motion in weakly nonlinear and dispersive media, and a suitable
approximation enables the KdV equation (for which wave propagation is
restricted to one direction) to be derived from this equation. Much work has
been reported for this equation, for example, the soliton and the
Painlev\'{e} expansion [16], the soliton and the bilinear form (Hirota's
method) [17], periodic wave solutions and the structure of the rational
solution to Boussinesq equation [18]$\ldots $, {\it etc}.

Here, the improved HB method is employed to obtain the singular solution of
the Boussinesq equation. Supposing the solution is of the form 
\begin{equation}
u=\sum_{i=0}^Nf^{(i)}(\omega (x,t)),  \label{5}
\end{equation}
where $N$ is an integer, substituting formula (5) into equation (3), and
balancing the nonlinear term $a(u^2)_{xx}$ and the linear term $bu_{xxxx}$,
we get $N=2$, and then

\begin{equation}
u=f^{^{\prime \prime }}\omega _x^2+f^{^{\prime }}\omega _{xx}.  \label{6}
\end{equation}
Substituting formula (6) into equation (3), and collecting all homogeneous
terms in partial derivatives of $\omega (x,t)$, we have

\begin{equation}
\begin{array}{c}
\lbrack bf^{(6)}-2af^{^{\prime \prime \prime }}f^{^{\prime \prime \prime
}}-2af^{^{\prime \prime }}f^{(4)}]\omega _x^6+[15bf^{(5)}-24af^{^{\prime
\prime }}f^{^{\prime \prime \prime }}-2af^{^{\prime }}f^{(4)}]\omega
_x^4\omega _{xx}+ \\ 
\lbrack f^{(4)}\omega _t^2\omega _x^2-f^{(4)}\omega
_x^4+(45bf^{(4)}-24af^{^{\prime \prime }}f^{^{\prime \prime
}}-12af^{^{\prime }}f^{^{\prime \prime \prime }})\omega _x^2\omega
_{xx}^2+(20bf^{(4)}- \\ 
8af^{^{\prime \prime }}f^{^{\prime \prime }}-4af^{^{\prime }}f^{^{\prime
\prime \prime }})\omega _x^3\omega _{xxx}]+[(\omega _{tt}\omega _x^2+\omega
_t\omega _x\omega _{xt}+\omega _t^2\omega _{xx}-6\omega _x^2\omega
_{xx})f^{^{\prime \prime \prime }}+ \\ 
(15bf^{^{\prime \prime \prime }}-6af^{^{\prime }}f^{^{\prime \prime
}})\omega _{xx}^3+(60bf^{^{\prime \prime \prime }}-20af^{^{\prime
}}f^{^{\prime \prime }})\omega _x\omega _{xx}\omega _{xxx}+(15bf^{^{\prime
\prime \prime }}-2af^{^{\prime }}f^{^{\prime \prime }})\omega _x^2\omega
_{xxxx}]+ \\ 
\lbrack (2\omega _{xt}^2+2\omega _x\omega _{xtt}+\omega _{tt}\omega
_{xx}-3\omega _{xx}^2+2\omega _t\omega _{xxt}-4\omega _x\omega
_{xxx})f^{^{\prime \prime }}+(10bf^{^{\prime \prime }}-2af^{^{\prime
}}f^{^{\prime }})\omega _{xxx}^2+ \\ 
(15bf^{^{\prime \prime }}-2af^{^{\prime }}f^{^{\prime }})\omega _{xx}\omega
_{xxxx}+6bf^{^{\prime \prime }}\omega _x\omega _{xxxxx}]+(\omega
_{xxtt}+b\omega _{xxxxxx}-\omega _{xxxx})f^{^{\prime }}=0.
\end{array}
\label{7}
\end{equation}
Setting the coefficient of $\omega _x^6$ in (7) to zero yields an ordinary
differential equation for $f$, namely

\begin{equation}
bf^{(6)}-2af^{^{\prime \prime \prime }}f^{^{\prime \prime \prime
}}-2af^{^{\prime \prime }}f^{(4)}=0.  \label{8}
\end{equation}
The solution of equation (8) is obtained as

\begin{equation}
f=-\frac{6b}a\ln \omega ,  \label{9}
\end{equation}
which yields

\begin{equation}
\begin{array}{l}
f^{^{\prime \prime }}f^{^{\prime \prime \prime }}=\frac b{2a}f^{(5)},\qquad
f^{^{\prime }}f^{(4)}=\frac{3b}{2a}f^{(5)},\qquad f^{^{\prime \prime
}}f^{^{\prime \prime }}=\frac baf^{(4)}, \\ 
f^{^{\prime }}f^{^{\prime \prime \prime }}=\frac{2b}af^{(4)},\qquad
f^{^{\prime }}f^{^{\prime \prime }}=\frac{3b}af^{^{\prime \prime \prime
}},\qquad \ \ f^{^{\prime }}f^{^{\prime }}=\frac{6b}af^{^{\prime \prime }}.
\end{array}
\label{10}
\end{equation}
Substituting formulae (10) into equation (7), it can be simplified to a
linear polynomial of $f^{\prime },f^{\prime \prime },\ldots $; then setting
the coefficients of $f^{\prime },f^{\prime \prime },\ldots $ to zero yields
a set of partial differential equations for $\omega (x,t)$,

\begin{equation}
\omega _t^2-\omega _x^2-3b\omega _{xx}^2+4b\omega _x\omega _{xxx}=0,
\label{11}
\end{equation}

\begin{equation}
\omega _{tt}\omega _x^2+4\omega _t\omega _x\omega _{xt}+\omega _t^2\omega
_{xx}-6\omega _x^2\omega _{xx}-3b\omega _{xx}^3+9b\omega _x^2\omega
_{xxxx}=0,  \label{12}
\end{equation}

\begin{equation}
2\omega _{xt}^2+2\omega _x\omega _{xtt}+\omega _{tt}\omega _{xx}-3\omega
_{xx}^2+2\omega _t\omega _{xxt}-4\omega _x\omega _{xxx}-2b\omega
_{xxx}^2+3b\omega _{xx}\omega _{xxxx}+6b\omega _x\omega _{xxxxx}=0,
\label{13}
\end{equation}

\begin{equation}
\omega _{xxtt}+b\omega _{xxxxxx}-\omega _{xxxx}=0.  \label{14}
\end{equation}
We note that the equation (14) from the coefficients of $f^{\prime }$ is a
linear PDE for $\omega (x,t)$. Then equations (11)-(13) can be regarded as
constraint conditions to the linear equation (14); thus solving equation (3)
reduces to solving the linear PDE (14) with some constraint conditions
(11)-(13).

Equation (14) may be integrated once to yield

\begin{equation}
\omega _{xtt}+b\omega _{xxxxx}-\omega _{xxx}=p(t),  \label{15}
\end{equation}
where $p(t)$ is an arbitrary function of time. It's easy to know that
equation (14) has the solution

\begin{equation}
\omega (x,t)=S(\xi )+q(t),  \label{16}
\end{equation}
where $S(\xi )$ is the traveling wave solution of equation $\omega
_{xtt}+b\omega _{xxxxx}-\omega _{xxx}=0$, and $q(t)$ satisfies equation $%
\frac{d^2}{dt^2}(q(t))=p(t)$. So the solution of equation (14) is of the form

\begin{equation}
\omega (x,t)=d_0+d_1(x-vt)+d_2(x-vt)^2+d_3e^{\beta (x-vt)}+d_4e^{-\beta
(x-vt)}+q(t),  \label{17}
\end{equation}
where $\beta =\sqrt{\frac{1-v^2}b}>0$; note that solution (17) is a
non-travelling wave solution of equation (14). Substituting formula (17)
into the constraint conditions (11)-(13), a set of ordinary differential
equations are obtained. Solving this set of equations, we find

\begin{eqnarray}
d_2 &=&0,  \label{18} \\
d_3 &=&\frac{bd_1^2(-1+4v^2)}{12d_4v^2(-1+v^2)},  \nonumber \\
q(t) &=&-\frac{d_1(-1+v^2)}vt.  \nonumber
\end{eqnarray}
Thus the solutions of equations (11)-(14) are obtained as

\begin{equation}
\omega (x,t)=d_0+d_1(x-vt)+\frac{bd_1^2(-1+4v^2)}{12d_4v^2(-1+v^2)}e^{\beta
(x-vt)}+d_4e^{-\beta (x-vt)}-\frac{d_1(-1+v^2)}vt,  \label{19}
\end{equation}
where $b$, $v$, $d_0$, $d_1$and $d_4$ are arbitrary constants. Substituting
solution (19) into formulae (9) and (6), the analytic solution of equation
(3) is obtained, namely

\begin{eqnarray}
&&u(x,t)=\frac{6b(d_1-d_4e^{-\beta (x-vt)}\beta +\frac{bd_1^2(-1+4v^2)\beta
e^{\beta (x-vt)}}{12d_4v^2(-1+v^2)})^2}{a(d_0+d_4e^{-\beta (x-vt)}-\frac{%
d_1(-1+v^2)t}v+\frac{bd_1^2(-1+4v^2)e^{\beta (x-vt)}}{12d_4v^2(-1+v^2)}%
+d_1(x-vt))^2}-  \label{20} \\
&&\frac{6b(\frac{d_4e^{-\beta (x-vt)}(1-v^2)}b+\frac{%
d_1^2(1-v^2)(-1+4v^2)e^{\beta (x-vt)}}{12d_4v^2(-1+v^2)})}{%
a(d_0+d_4e^{-\beta (x-vt)}-\frac{d_1(-1+v^2)t}v+\frac{bd_1^2(-1+4v^2)e^{%
\beta (x-vt)}}{12d_4v^2(-1+v^2)}+d_1(x-vt))},  \nonumber
\end{eqnarray}
where $a$, $b$, $v$, $d_0$, $d_1$and $d_4$ are arbitrary constants. Here $a$
just is a parameter to decide the relative value of $u(x,t)$, and when $%
d_1=0 $, the solution (20) decays to the solitary wave solution.

Solution (20) is very complex and includes six arbitrary constants, so it
needs discussion. It's easy to understand that the singularity of solution
(20) comes form the zero value of formula (17), namely, it is decided by $%
\omega (x,t)=0$. If the analytic solution $x=T(t)$ or $t=X(x)$ of $\omega
(x,t)=0$ is obtainable, the main property of the solution can be given, but
formula (17) includes the transcendental function, so it's impossible to get
an explicit solution. By numerical analyticity, the main property of the
solution is reported as follows: when $0.5>v\ $or$\ v<-0.5$, $b<0$, $d_1<0$, 
$d_4<0$, the unlimited-time blow-up solution exists, and when $0.5\geq v\geq
-0.5(v\neq 0)$, $b>0$, $d_1>0$, $d_4>0$, the finite-time blow-up solution
exists. Fig.1. shows a finite-time blow-up solution. The numerical results
suggest that: (1) the constant $v$ decides the region of the singularity,
when $0.5\geq v>0$, the singularity exists in the half-line $[-\infty ,t_0]$
and when $0>v\geq -0.5$, the singularity exists in the half-line $%
[t_0,\infty ]$; (2) the constant $a$ just decides the relative values of $%
u(x,t)$, and the constants $b$, $d_0$, $d_1$, $d_4$ decide the values of $%
t_0 $, which is the blow-up time; and (3) there are two kinds of finite-time
blow-up evolution modes at $v=-0.5$ or $v=0.5$ and $0.5>v>-0.5$. Fig.2. ($%
v=-0.5$) shows a time evolution mode of the finite-time blow-up solution.
Fig.3. ($0>v>-0.5$) shows another time evolution mode of the finite-time
blow-up solution.

\section{KdV equation and gKdV equation}

As is known, the KdV equation is not only of mathematical interest but also
of practical importance. It has been shown to describe small amplitude
shallow water waves, hydromagnetic waves in a cold plasma, ion-acoustic
waves, acoustic waves in an anharmonic crystal, and wave motions in other
biological and physical systems. The standard KdV equation is given by

\begin{equation}
u_t+\alpha uu_x+\beta u_{xxx}=0,  \label{21}
\end{equation}
where $\alpha $, $\beta $ are real constants. When the improved HB method is
applied to the KdV equation, the following analytic solution of KdV equation
is obtained

\begin{equation}
\begin{array}{l}
u(x,t)=(48d_4e^{b(x-vt)}v(3\beta ^2d1^3e^{2b(x-vt)}b+\beta
d_1v(-d_1e^{b(x-vt)}(8d_4+ \\ 
d_0e^{b(x-vt)})+8d_4^2b+d_1^2e^{2b(x-vt)}(3vt-x))+4d_4^2v^2(d_0+d_1(x-3vt))))
\\ 
/(a(\beta d_1^2e^{2b(x-vt)}-4d_4v(d_4+e^{b(x-vt)}(d_0+d_1(x-3vt))))^2).
\end{array}
\label{22}
\end{equation}
It is easy to show that solution (22) is an unlimited-time singular solution.

Recently, some generalized KdV equations (gKdV) have been discussed widely,
including a few high order KdV equations [19], the q-KdV equation [20], and
some generalized KdV equations [21]. One generalized KdV [10] 
\begin{equation}
u_t+(n+1)(n+2)u^nu_x+u_{xxx}=0,  \label{23}
\end{equation}
has been presented and a more generalized form 
\begin{equation}
u_t+a(u^m)_x+b(u^l)_{xxx}=0,  \label{24}
\end{equation}
has been introduced by Rosenau and Hyman [22]. Equation (24) has yielded the
compactons solution [23] for certain values of $m$ and $l$, which, like
solitons, have the remarkable property that after colliding with other
compactons, they reemerge in the same coherent shape. In the paper, a
concrete equation of the gKdV equation (23) (in the case $n=3$)

\begin{equation}
u_t+20u^3u_x+u_{xxx}=0.  \label{25}
\end{equation}
is considered. For a discussion of equation (25), the nonlinear
transformation

\begin{equation}
u(x,t)=\frac{g(x,t)}{f(x,t)^{\frac 23}}+u_0(x,t),  \label{26}
\end{equation}
is introduced. Then, substituting solution form (26) into equation (25), and
making the coefficients of like powers of $f(x,t)$ vanish, we obtain the
following set of equations,

\begin{equation}
\begin{array}{l}
f^{-\frac{11}3}:-\frac{40}{27}gf_x(9g^3+2f_x^2)=0; \\ 
f^{-\frac 93}:-40g^3u_0f_x=0; \\ 
f^{-\frac 83}:\frac{10}3(6g^3g_x+f_x^2g_x+gf_xf_{xx})=0; \\ 
f^{-\frac 73}:-40g^2u_0^2f_x=0; \\ 
f^{-\frac 63}:20g^2(3u_0g_x+gu_{0x})=0; \\ 
f^{-\frac 53}:-\frac 23(gf_t+20gu_0^3f_x+3g_xf_{xx}+3f_xg_{xx}+gf_{xxx})=0;
\\ 
f^{-\frac 43}:60gu_0(u_0g_x+gu_{0x})=0; \\ 
f^{-\frac 23}:g_t+20u_0^3g_x+60gu_0^2u_{0x}+g_{xxx}=0,
\end{array}
\label{27}
\end{equation}
where $u_0(x,t)$ satisfies the equation

\begin{equation}
u_{0t}+20u_0^3u_{0x}+u_{0xxx}=0,  \label{28}
\end{equation}
which is the same as equation (25) for $u(x,t)$; thus the set of equations
(26-28) constitutes an invariant-B\"{a}cklund transformation of the g-KdV
equation (25).

It is obvious that equation (28) has the trivial solution $u_0(x,t)=0$, then
substituting $u_0(x,t)=0$ into formula (26), the transformation is simply

\begin{equation}
u=\frac{g(x,t)}{f(x,t)^{\frac 23}},  \label{29}
\end{equation}
and equations (27) become

\begin{equation}
\begin{array}{l}
g_t+g_{xxx}=0; \\ 
gf_t+3g_xf_{xx}+3f_xg_{xx}+gf_{xxx}=0; \\ 
6g^3g_x+f_x^2g_x+gf_xf_{xx}=0; \\ 
9g^3+2f_x^2=0.
\end{array}
\label{30}
\end{equation}
The frist and last equations are easily solved (the other equations can be
treated as constraint conditions), from which solutions of equations (30)
can be obtained. Substituting the solutions into transformation (29), the
analytic solution of the g-KdV equation(25) is obtained as

\begin{equation}
u(x,t)=-\frac{v^{\frac 13}\sec ^{\frac 43}(\frac{-16c+9v(x+vt)}{12\sqrt{v}})%
}{2\tan ^{\frac 23}(\frac{-16c+9v(x+vt)}{12\sqrt{v}})},  \label{31}
\end{equation}
where $c$ is a constant, and $v$ is the velocity of the travelling wave.
It's easy to understand that the singularity of solution (31) is decided by
equation $\tan ^{\frac 23}(\frac{-16c+9v(x+vt)}{12\sqrt{v}})=0$. Thus the
solution (31) is a space periodic singular solution, {\it i.e.,} the g-KdV
equation contains the unlimited-time singular solution, which is periodic in
space.

\section{Conclusion}

In conclusion, two methods to seek analytic solutions of nonlinear PDEs are
introduced: an improved HB method and the invariant-B\"{a}cklund
transformation based on a special nonlinear transformation, and two kinds of
analytic singular solutions (finite-time and infinite-time singular
solutions) of the Boussinesq equation and a g-KdV equation (including the
KdV equation) are obtained. A finite-time singular solution of the
Boussinesq equation is discussed, which is relevant to singularity formation
in finite time in hydrodynamics. A periodic in space unlimited-time singular
solution of the g-KdV equation is obtained, and the invariant-B\"{a}cklund
transformation of the equation is presented. In addition, the nonlinear
transformation $u=\frac g{f^{\frac 2n}}+u_0$ is useful in finding solutions
of equation (23).

{\bf Acknowledgments: }This work was supported by the National Natural
Science Foundation of China.

The corresponding email address: yangkq@lzu.edu.cn (Yang Kongqing).

{\bf Fig. Caption: }Quantities plotted in the figures are dimensionless.

Fig.1. A finite-time blow-up evolution is shown of a finite-time singular
solution; the values of the parameters are as follows: $a=-0.1$, $b=1$, $%
v=-0.5$, $d_0=0.5$, $d_1=1$, $d_2=1$. In the figure, the max value $u(x,t)$
and the asymptotic blow-up time is indicated.

Fig.2. A finite-time blow-up solution is shown; the values of the parameters
are as follows: $a=-0.1$, $b=1$, $v=-0.5$, $d_0=0.5$, $d_1=1$, $d_2=1$. At $%
t=0$, $u(x,t)$ looks like a hump(a standard solitary wave), the height of
the hump becomes higher and higher, and at $t\simeq 1.0$ the height of the
hump tends to infinity.

Fig.3. A finite-time blow-up solution is shown; the values of the parameters
aret as follows: $a=-0.1$, $b=1$, $v=-0.2$, $d_0=0.5$, $d_1=1$, $d_2=1$. At $%
t=0$, $u(x,t)$ looks like two humps, one hump runs to another hump, at $%
t\simeq 1.0$, two humps come into collision and become a single hump, after
which the height of the hump tends to infinity.

\end{document}